Methods note /

# On Solving Groundwater Flow and Transport Models with Algebraic Multigrid Preconditioning

M. Adil Sbai[1] and A. Larabi[2]

[1] Corresponding author: BRGM, Water, Environment, Process and Laboratories Division (DEPA), Water Management Group, 3, Avenue Claude-Guillemin, BP 36009, 45060 Orléans Cedex 2, France; Fax: +33238643719; Tel: +33238643527; sbai.adil@gmail.com; a.sbai@brgm.fr

[2] Regional Water Centre of Maghreb, LIMEN, Ecole Mohammadia d'Ingénieurs, Université Mohammed V in Rabat, B.P. 765 Agdal Rabat, Morocco; larabi@emi.ac.ma

## Abstract

Iterative solvers preconditioned with algebraic multigrid have been devised as an optimal technology to speed up the response of large sparse linear systems. In this work, this technique was implemented in the framework of the dual delineation approach. This involves a single groundwater flow solve and a pure advective transport solve with different right-hand sides. The new solver was compared with traditional preconditioned iterative methods and direct sparse solvers on several two- and three-dimensional benchmark problems spanning homogeneous and heterogeneous formations. For the groundwater flow problems, using the algebraic multigrid preconditioning speeds up the numerical solution by one to two orders of magnitude. Contrarily, a sparse direct solver was the most efficient for the pure advective transport processes such as the forward travel time simulations. Hence, the best sparse solver for the more general advection-dispersion transport equation is likely to be Péclet number dependent. When equipped with the best solvers, processing multimillion grid blocks by the dual delineation approach is a matter of seconds. This paves the way for routine time-consuming tasks such as sensitivity analysis. The paper gives practical hints on the strategies and conditions under which algebraic multigrid preconditioning for the class of nonlinear and/or transient problems would remain competitive.

**Introduction**

Groundwater models become a fundamental component of the sustainable management of water resources. This trend is accelerating at a faster rate during the last decades. As Moore's law is approaching definitive obsolescence, computational demands for groundwater models are still rising. Runtimes can be reduced by adopting novel hardware technologies, fast solvers, multiscale methods or reduced-order techniques, just to cite a few. While some of these strategies are at an early stage of developments, the latest generation of sparse iterative and/or direct methods in the linear algebra field are mature enough to support some of these needs (Saad, 2003; Davis 2006). Indeed, one of the greatest advances in groundwater modeling technology during the last decade is the adoption of algebraic multigrid methods (AMG) that efficiently solve large sparse linear systems of discrete equations. This is going mainstream as many vendors of widely used industrial groundwater modeling software offer AMG as an alternative solver and/or preconditioning method out of the box. However, these software packages give little if no guidelines on the optimal use of the AMG for groundwater applications implying further research. For instance, the SAMG library developed by the Fraunhofer Institute for Algorithms and Scientific Computing (Fraunhofer SCAI 2020) was integrated into FEFLOW (Diersch 2013) and some MODFLOW versions (AquaVeo 2020; Waterloo Hydrogeologic 2020). There exist many other closed- and open-source implementations of the AMG method. However, plugging them into existing research and/or in-house groundwater computer programs is often challenging due to significant investment costs. Many institutions or small research groups cannot afford these additional efforts. For instance, specific implementations of AMG are offered by the well-known PETSc and Trilinos libraries. These are huge scientific computing toolkits, which are not easily portable to non-Unix-like operating systems. For most research projects, this heavy machinery is an excessively large and unnecessary dependency. Moreover, some libraries assume a predefined sparse matrix storage format and/or parallelization technology that unsuits the target groundwater modeling code. Dedimov (2019) describes these monolithic approaches with pointers to other AMG implementations around. This author developed the AMGCL library (AMGCL 2020), as an alternative, with such practical constraints in mind. This open-source library was released under a liberal MIT license permitting its integration into proprietary or commercial codes.

Development of the multigrid methods started since the 1980s of the last century (Stüben 2001). Within simple iterative algorithms, the convergence rate is slow for smooth waves and fast for oscillatory waves. In the framework of multigrid methods, a linear system is solved on a hierarchy of increasingly coarse grids, such that the long and short-range Fourier mode frequencies are decomposed leading to a faster convergence (Briggs et al. 2000). The initial efforts went into explicit definition of these grids as advocated by the geometric multigrid methods (GMG). As indicated by Detwiler et al. (2002) and references cited therein, GMG has attracted little applications in groundwater modeling practice due to its inherent limitations to adapt to complex geometries and the need to manually specify the hierarchical grids. Algebraic multigrid methods automate this process leading to a 'black-box' solver. Indeed, AMG methods work directly on the algebraic level by a nodal aggregation to recursively define coarser linear systems.

The algorithm starts with a setup phase where the grid hierarchy is determined automatically. During this phase, prolongation and restriction operators that transfer information between the grids at different levels are equally determined. This setup phase is the most difficult from an algorithmic perspective, but usually takes an insignificant fraction of the overall AMG computational effort.

In the second phase, the algorithm performs a number of sweeps along the so-called V-cycle. Starting from an initial guess, the residual of the linear system is restricted onto the next coarse level. Then, a smoothing step is performed followed by the residual evaluation. This loop continues until reaching the coarsest level where a direct solver is used to compute a small linear system. Next, the computed vector of unknowns is interpolated into the upper fine level, used to correct the residuals, and to smooth the linear system. The process of interpolating and smoothing the residuals continues until reaching the top. Moving downwards and upwards along the V-cycle involve sparse matrix-vector products only, making the algorithm very fast. More details on the AMG methods may be found in (Ruge and Stüben 1987; Briggs 2000; Stüben 2001).

While the algebraic multigrid method is a sparse iterative solver by its own, it can be used as a preconditioner to accelerate the convergence rate of iterative Krylov subspace methods (Saad 2003). This is the approach followed by the handful set of the previously reported applications involving steady-state groundwater flow problems. Ashby and Falgout (1996) reported the first related work in the framework of the ParFlow hydrologic model (ParFlow 2020). Detwiler et al. (2002) have compared the performances of the MODFLOW built-in solver PCG2 and that of the legacy AMG1R5 solver after its linkage with MODFLOW. Kourakos and Harter (2014) have employed an AMG preconditioner to produce a highly detailed velocity field at a groundwater basin scale in California. Their model was used for non-point source pollution in combination with the streamline simulation approach to analyze nitrate breakthrough at a large number of groundwater receptors. They have compared the performances of many variants of the AMG preconditioners provided by the Trilinos and HYPRE libraries. Serial implementations have reported speedups ranging between one and two orders of magnitude for steady-state groundwater flow problems (Detwiler et al. 2002; Thum and Stüben 2012).

The only study that showed the use of AMG preconditioning for a solute transport problem was that by Detwiler et al. (2002) where the ORTHOMIN accelerator (Saad 2003) was chosen to solve a steady-state transport problem. However, solving steady-state solute transport problems is of little interest to practitioners. Nevertheless, scalar steady-state transport like equations were recently proposed in the framework of the dual delineation approach. This method first introduced by Sbai (2018) and extended in (Sbai 2019) solves a total of $N_i+N_P+2$ stationary scalar advection equations to calculate travel times, steady-state capture zones and other derived quantities such as the time-related capture zones, etc. $N_i$ and $N_p$ are the number of injection (and inflowing boundaries) and pumping (and outflowing boundaries) wells, respectively. Moreover, the underlying sparse linear systems have the same matrix enabling to reuse its factorization during the course of direct or iterative methods.

Therefore, large-scale transport problems particularly associated with this method are ideally suited for AMG preconditioning and will be used as typical benchmarks in this work.

In the framework of an IMPES (IMplicit Pressure Explicit Saturation) formulation (Chen et al. 2006) for the immiscible two-phase flow problem, Wendland and Flensberg (2005) reported the higher efficiency of AMG preconditioning, when solving the inner linear systems, than traditional preconditioners. Although this was demonstrated only on two-dimensional problems, the approach would be important to pursue for large-scale three-dimensional models.

The objective of this research is two-fold. First, we evaluate the suitability of the AMGCL library for groundwater modeling applications. Second, we compare the performances of state-of the-art sparse direct and preconditioned iterative solvers for two- and three-dimensional groundwater models based on the dual delineation approach. We start by a short description of general features and the structure of the AMGCL library. Next, we briefly describe groundwater flow and transport models selected as test beds to benchmark the library performances before commenting on the obtained results. Then we give practical hints to extend the method efficiency for the class of nonlinear and/or transient groundwater problems indicating areas for future research. Finally, a concluding remarks section closes the paper.

**The AMGCL Library**

AMGCL is a header-only, open-source, C++11 library implementing the algebraic multigrid method. The numerical solution step of the AMG may use various parallelization technologies, such as OpenMP, OpenCL and CUDA. The adopted generic object-oriented design facilitates transparent plugging of the library into existing scientific computing codes. The library supports different combinations of linear iterative solvers for sparse symmetric and non-symmetric matrices, coarsening strategies and relaxation methods.

It is also possible to call the library from other object-oriented languages such as, Python, FORTRAN 2003, and Object Pascal (i.e. from the Delphi IDE). A FORTRAN module using the intrinsic `iso_c_binding` module provides interfaces of callable AMGCL routines from external programs. This module can facilitate, therefore, the integration of AMGCL into many legacy groundwater codes primarily developed with this programming language. Installation for use from the supported computer languages is performed with the included CMake scripts. Several examples demonstrating how to use the library are included in the distribution.

Besides these out of the box bindings, the computational geosciences group in the Department of Mathematics and Cybernetics at SINTEF Digital developed a MATLAB$^©$ exchange interface to the MATLAB Reservoir Simulation Toolbox (Lie 2019). This interface provided a convenient means to upgrade an existing MATLAB-based groundwater flow and transport modeling toolbox (Sbai 2018; 2019). All benchmarks presented in this note were performed within this framework by using simple MATLAB scripts where the groundwater models were simultaneously pre-processed, executed and post-processed.

**Benchmarking Examples**

We present two test problems to evaluate the performance of the AMGCL library. The first test problem is a two-dimensional field-scale application that was published in previous works (Sbai 2018; 2019) demonstrating the so-called dual delineation technique. It involves four well doublets in a deep geothermal reservoir where heat production from pumping wells and cold fluid injection are taking place simultaneously. All model boundaries are no-flow Neumann type. A uniform grid spacing of 20m in all directions was considered leading to 275,000 grid blocks.

The second test problem is a three-dimensional conceptual extension of the first model. The total aquifer thickness was uniformly divided into 10 layers such that the problem size is one order of magnitude higher. Injection and production wells have partial penetrations along the five lower and top layers, respectively. This creates a three-dimensional flow field because the vertical velocity components in the wells neighborhoods are expected to be significant.

For each problem, two distributions of the hydraulic conductivity tensor, **K**, were considered. At first, a homogeneous and isotropic hydraulic conductivity that equals $K_{mean}$ = 6.38 $10^{-5}$ m/s was taken. Next, two synthetic 2D and 3D lognormal random **K** fields were generated with mean hydraulic conductivities $K_{mean}$ and variance of $\sigma_{lnK}^2 = 10$, and assigned to each test problem respectively. Because standard 5-point and 7-point stencils result from finite-difference discretization of the groundwater flow equation, the resulting sparse systems matrices are M-matrices in all cases. This is a suitable property guarantying the convergence of iterative conjugate gradient (CG) methods (Saad 2003). Note that groundwater finite element models based on the Galerkin weighted residual approach (Huyakorn and Pinder 1983; Wang and Anderson 1995) do not lead natively to such matrices when distorted elements, such as irregular quadrilaterals of hexahedra, are used. To correct this undesirable behavior, a physically based M-matrix transformation of the global conductance matrix, enforcing local mass conservation, could be ideally applied prior to the preconditioning step (Larabi and De Smedt 1994). A similar issue occurs for the conforming finite element method when applied to solute advection-dispersion transport problems. An appropriate M-matrix transformation has shown to be equally effective in such case (De Smedt and Sbai 1998).

The simulations were divided into two groups belonging to the stationary groundwater flow and forward travel time simulations for each test case.

For the groundwater flow simulations, we compare results computed with AMGCL solvers and those obtained with two other solvers. The first is the built-in MATLAB's PCG with an incomplete Cholesky factorization preconditioner and denoted by CG/IC(0) in the following. This can be regarded as an equivalent of the legacy PCG2 package distributed within MODFLOW. Next, the built-in MATLAB's symmetric direct solver based on the sparse supernodal Cholesky algorithm of the CHOLMOD library was selected (Chen et al. 2008; Davis 2006).

For the forward travel time simulations, the performances of AMGCL solvers are compared with those obtained with the BiCGSTAB (Van der Vorst 1992) solver provided by MATLAB with ILU(0) preconditioning and the UMFPACK direct sparse solver based on the unsymmetrical multifrontal algorithm (Davis 2004; 2006). The solved systems, subject to Dirichlet boundary conditions, are given as

$$+\vec{v} \cdot \nabla \tau_f = \phi \quad \tau_f|_I = 0 \qquad (1)$$

where, $\vec{v}$ [LT$^{-1}$], is the Darcy velocity, $\tau_f$ [T] is the forward travel time; $\phi$ [-] is the aquifer porosity; and $I$ is the set of all injection wells. Because solving Equation 1 is quite novel in groundwater modeling (Sbai 2018; 2019) selection of the most efficient solver/preconditioner is of great interest to enhance the competitiveness of this method. Unlike in (Sbai 2018) where a multistep time integration approach was proposed to solve Equation 1 as a shortcut to recycle existing transient transport models, a true steady-state one-step solver was used in this work.

The tolerance for the convergence of the linear sparse iterative solvers has been set at $10^{-12}$ for all the presented test problems. Notably, direct solvers feature no user control parameters.

## Results

All reported simulations were executed on a DELL Latitude 3520 series laptop equipped with 16 GB of 2400MHz DDR4 RAM and an Intel(R) core(TM) i5-7300HQ 2.50GHz CPU. We have used the MATLAB R2018a release for Microsoft Windows 10 operating system.

### *Groundwater Flow Problems*

Table 1 shows an example of the reported output of the AMGCL based solvers for a groundwater flow problem. It shows, in particular, that the first phase for the solver setup time takes only a few seconds even for the largest models. For the homogeneous three-dimensional problem, there were five hierarchical levels going from the finest level, denoted by 0, to the coarsest one where there were only 589 unknowns to solve. AMGCL reports also the random access memory (RAM) used to store the linear system at each level and how this amounts to the total used memory for storage.

Table 1 – Example of the AMGCL solver output for the homogeneous three-dimensional test problem. General information on the used solver and preconditioner along with their memory footprint are given. The computed

grid hierarchy shows the number of linear system unknowns, sparse matrix non-zeros and the memory taken for RAM storage at each level.

```
Solver setup took 2.943 seconds
Solver
======
Type:             CG
Unknowns:         2750000
Memory footprint: 83.92 M

Preconditioner
==============
Number of levels:    5
Operator complexity: 1.93
Grid complexity:     1.45
Memory footprint:    1.51 G

level     unknowns       nonzeros        memory
---------------------------------------------------
    0      2750000       18679000     838.59 M (51.71%)
    1      1100000       13729000     580.69 M (38.01%)
    2       137668        3604176     127.59 M ( 9.98%)
    3         5208          96230       3.61 M ( 0.27%)
    4          589          11797     391.69 K ( 0.03%)
```

Figure 1 compares the performances of three AMGCL linear solvers with those corresponding to the CHOLMOD sparse direct solver and the standard conjugate gradient solver preconditioned with an incomplete Cholesky factorization. Unsurprisingly, the direct solver is generally the winner for 2D problems whereas it is the worst choice for 3D problems. CG/ILU(0) and CG/IC(0) solvers are of comparable performance for all tested problems. They are the slowest for 2D problems and much faster than the direct solver for 3D problems. Using AMG as a preconditioner is highly effective in all cases with a speedup ranging between one and two orders of magnitude when taking CG/IC(0) as a reference. This is in close agreement with the range of previously reported speedups in previous works (Detwiler et al. 2002; Thum and Stüben 2012). The most important speedups were obtained for the 3D problems. The smoothed aggregation scheme for AMG coarsening slightly enhances the solution efficiency for case studies with anisotropic and heterogeneous subsurface properties. Similarly, tuning the relaxation method affects the performance of AMG preconditioners. In this work, ILU(0) relaxation was used throughout because it was found to be more efficient than other schemes, such as Gauss-Seidel and damped Jacobi.

The solver performance hierarchy was overall maintained when moving from homogeneous to heterogeneous subsurface realizations for each test problem.

Figure 2 compares the memory usage by the considered solvers for the homogeneous three-dimensional test problem. The direct solver had the highest storage demands. PCG methods preconditioned with AMG are lying at an intermediate category where RAM usage is less than for the direct solver but is still significant. The classical CG/IC(0) and CG/ILU(0) methods had the lowest computer storage requirements explaining the success of these PCG methods during the 1980s-1990s era when computational resources at that time were much modest than today's standards. In view of this, the additional cost of higher memory storage when moving towards AMG preconditioning is affordable, such that the related assertions by Detwiler et al. (2002) are no longer of concern.

*Transport Problems*

Figure 3 compares the performances of two AMGCL linear solvers with those obtained by the direct sparse solver UMFPACK and the BiCGSTAB accelerator preconditioned with an incomplete factorization. The direct solver is the most efficient for all test problems. It largely outperforms all sparse iterative solvers for the three-dimensional test cases. This is, to our best knowledge, an unreported result in the current groundwater modeling literature where preconditioned iterative methods are also the norm for transport problems. This also indicates that the linear solvers implemented in many widely used solute/heat transport packages, such as MT3DMS (i.e. ORTHOMIN), are becoming outdated for a range of subsurface applications and need to be updated accordingly. The widely used BiCGSTAB /ILU(0) solver for transport problems is the least efficient for all tested problems. Using AMG as a preconditioner is a better option than ILU(0) as it is generally two times faster.

Contrarily to flow problems, the solution efficiency was found to be less sensitive to the coarsening and relaxation schemes. However, the solver performance hierarchy remains identical when moving from homogeneous to heterogeneous subsurface test problems.

Figure 4 compares the RAM usage by the considered solvers for the homogeneous three-dimensional test problem. AMG preconditioning is the most RAM consuming method. The direct solver is not only efficient but had lower storage demands which are comparable to the legacy BiCGSTAB/ILU(0) solver.

Notably, there exists a large choice among sparse direct methods for unsymmetrical matrices available in many software packages (Davis 2006). Some algorithms have multithreaded or massively parallel versions, such as for SuperLU. Therefore, other algorithms may give even superior efficiency.

Figure 5 compares the speed up of different solvers for the flow and transport sub-problems of the 3D test cases. Reported values scale the CPU times with that for the least efficient solver, excluding the direct solver for the flow problems. It is clear from Figure 5a that the relative speedup of AMG preconditioning decreases with heterogeneity. There was no similar impact on the performance of the transport solvers (Figure 5b).

**Discussion**

Based on the obtained results we recommend the use of different types of sparse solvers in the framework of the dual delineation approach. The iterative CG/AMG and direct UMFPACK solvers are the best choices for the flow and transport problems, respectively. While CG/AMG is an ideal choice for the steady state and confined (i.e. linear) case, its performance is not easily maintainable in the most general cases. In the following, we discuss turnarounds and possible future extensions for typical groundwater modeling problems.

For steady-state unconfined or variably saturated groundwater flow problems, it was argued that the efficiency of AMG preconditioning could be preserved by caching its setup phase (Thum and Stüben 2012), so it was done only once at the beginning of the nonlinear iteration loop. This strategy is only effective for slowly changing coefficients of the sparse conductance matrix. This is typically the case for modified Picard iteration schemes owing to their slow convergence rate. Within a Newton-Raphson iteration, this is unlikely because there is no guarantee that the Jacobian matrix entries will maintain a slow variation. Under unsaturated conditions, this will strongly depend on the type of soils characteristic curves retained in the model. Overall, the nonlinear iteration scheme may shadow the competitiveness of the AMG preconditioning. The extent of this assertion, however, deserves thorough investigations.

For transient flow problems, it is unlikely that one preconditioner will remain the most efficient during many stress periods with contrasting anthropogenic and hydrological forcing. Thum and Stüben (2012) suggested a dynamic switch between many preconditioning methods based on previously recorded memory requirements, runtime and convergence rate. However, these authors did not explain the technique to a sufficient detail enabling reproducibility. They concluded that AMG preconditioning should be selected whenever large time steps are used. This is, however, not the only criterion for transient problems because aquifer storativity and grid size distributions are equally important in this context.

Let us consider the standard fully implicit finite difference in time approximation of the transient groundwater flow equation. This reads in matrix form (Huyakorn and Pinder 1983; Wang and Anderson 1995).

$$\left(\mathbf{C} + \frac{\mathbf{S}}{\Delta t^{n+1}}\right) h^{n+1} = Q^{n+1} \qquad (2)$$

where $\mathbf{C}$ is the global conductance matrix resulting from a given spatial discretization method such finite differences (FDMs) or the conforming finite elements (FEMs). $\mathbf{S}$ is a diagonal storativity matrix whose entries are $S_i V_i$. $V_i$ is the volume of cell i or the control-volume of the i$^{th}$ node when considering the FDMs and FEMs methods, respectively. $S_i$ is the specific storage coefficient or the specific yield of the aquifer volume at cell i. $\Delta t^{n+1}$ is the time step at the current time level $n + 1$. $h^{n+1}$ are unknown groundwater heads, and $Q^{n+1}$ is a flow rate vector holding the contributions from prescribed boundary conditions at this time level and from groundwater heads in the previous time step (i.e. $h^n$).

Hence, it becomes clear that one can select an AMG preconditioner and keeps caching the AMG setup phase when $\frac{S_i V_i}{\Delta t^{n+1}} < \delta C_{ii}$ is fulfilled for all diagonal entries, where $\delta$ is a small scaling factor (i.e. 0.01). Thus, the condition of sufficiently large time steps as suggested by Thum and Stüben (2012) is not sufficient. This implies that transient problems with either coarser grids or soil materials having large specific yields, such as coarse granular sands, are not naturally suited to AMG preconditioning. Therefore, soil material properties, spatial and temporal discretizations are parameters that rule the preconditioning choice for transient groundwater flow problems.

Following similar reasoning, reusing an AMG preconditioner during a transient solute/heat transport simulation becomes possible when the entries of the mass matrix divided by the current time step are relatively small relative to contributions from the advection and hydrodynamic dispersion operators. Moreover, we expect different behavior of the solvers for problems with different Péclet regimes. As demonstrated above, sparse direct solvers may be preferred for advection-dominant problems. For dispersion-dominant problems, a solute adsorbing on the rock surface will be less favorable to AMG preconditioning than a conservative tracer. This is because diagonal matrix entries for a sorbing solute are much higher than that associated with a tracer. Implementations using time splitting approaches can benefit from the contrasted performances of the sparse direct and preconditioned iterative solvers for the advection and dispersion operators.

Additionally, many groundwater management problems involve coupling between the flow and the transport solvers. These are notably more challenging problems needing advanced computational methods. Examples are density- and viscosity- dependent coupled flow and transport formulations targeting applications such as seawater intrusion in coastal aquifers, subsurface heat extraction and storage, among others (Huyakorn and Pinder 1983). A bottleneck of the underlying solvers is the high computational cost involved when repeatedly solving the pressure equation to update the flow field. The same issue also occurs in two-phase flow problems based on a sequential implicit pressure-saturation formulation (Chen et al. 2006). The constrained pressure residual (CPR) preconditioner recently implemented in AMGCL for similar fully implicit coupled problems encountered in reservoir engineering (Gries et al. 2014) is an interesting method to explore in the future.

This technique applies an AMG preconditioning solely on the block associated to the pressure unknowns in the globally assembled system matrix.

While the authors have enjoyed using AMGCL to solve large-scale steady-state groundwater problems in their laptops, they believe they are only scratching the surface of what is being possible to do with this library for groundwater modeling. The AMGCL design enables a transparent port of legacy groundwater codes into a diversity of modern high performance computing platforms such as distributed systems and GPGPUs: a promising path for future groundwater modeling applications.

Our overall impression is that AMGCL is a stable, flexible, and credible AMG implementation. It enables tight integration of the AMG technology with legacy groundwater modeling codes. A task that we already started for other in-house research codes. The provided tutorials in C++ are clear and easy to follow. However, the learning path to master the provided solver options needs an advanced background in AMG methods. This knowledge is essential to tune up the AMGCL solver for a particular groundwater application; meanwhile the promise of using a black-box solver library remains valid. The AMGCL project could provide better documentation of the examples and on how to use the Python and FORTRAN wrappers. All in all, AMGCL is a well-designed and powerful lightweight AMG library with a great potential to empower research and professional subsurface hydrology simulations. Its liberal license enables its unrestricted integration with the widely used groundwater flow and transport codes such as MODFLOW-USG, MODFLOW6 and MT3DMS. This will be of great benefit to the hydrogeological modeling community worldwide.

Finally, we shall bear in mind that the algebraic multigrid method was designed to solve large-scale stationary sparse linear systems. While the above-discussed tips and tricks to extend the method's efficiency for some nonlinear and/or transient problems might work, this is a counter nature strategy. Emerging time integration with multigrid approaches that apply the multigrid technique to the time dimension are promising alternatives (Falgout et al. 2014). The multigrid reduction in time (MGRIT) is supported by the non-intrusive XBraid software package developed by the LLNL (XBraid 2020). However, these are novel computational methods, which are under constant developments. They have been showed to be effective for hyperbolic problems similar to transient groundwater flow. This is not the case for all transport problems occurring in subsurface aquifers.

**Concluding Remarks**

This paper compares the performances of many preconditioned sparse iterative solvers implemented in the AMGCL library and selected state-of-the-art sparse direct solvers for steady-state groundwater flow and transport such as the grid-based travel time simulations. The following general conclusions can be drawn from this study.

1. The obtained results are in agreement with the previously reported speedups when using the algebraic multigrid as a preconditioner for the steady-state groundwater flow. Obtained speedups range between one and two orders of magnitude.
2. The sparse direct solver, UMFPACK, was the most efficient solver for unsymmetric systems arising from the pure advection processes such as travel times. For 3D problems, it largely supersedes the more traditional Krylov iterative methods with incomplete factorization preconditioning.
3. The dual delineation method equipped with this couple of efficient solvers was able to process problems with multimillion grid blocks in a handful set of seconds on commodity PC hardware.

4. AMGCL testing on the selected benchmarks went smoothly while its integration required only a few lines of code. The library has a great potential to empower research and professional subsurface hydrology simulations

Practical hints on future research pathways towards a quantitative assessment of the most efficient sparse solvers for nonlinear and/or transient groundwater models are given. Smarter engines having the predictive ability to switch to the most efficient solver (i.e. direct vs iterative) and/or preconditioner during the course of a single simulation need to be developed. Such computational strategies are not yet available at this time and are therefore subject to further research.

# Figures Captions

Figure 1 – Performance of the direct and AMGCL solvers for the (a) two- and (b) three-dimensional groundwater flow test problems. CPU times of the linear solvers are reported in logarithmic axis scale. (CHOLMOD: Direct sparse solver from MATLAB; CG/IC(0): Conjugate gradient solver preconditioned with Incomplete Cholesky factorization from MATLAB; CG/ILU(0): Conjugate gradient solver preconditioned with ILU(0) from AMGCL; CG/AMG/RS: Conjugate gradient solver preconditioned with AMG using Ruge-Stüben coarsening; CG/AMG/SA: Conjugate gradient solver preconditioned with AMG using smoothed aggregation coarsening).

Figure 2 – Comparison of the memory usage by the tested groundwater flow solvers for the homogeneous three-dimensional benchmark problem.

Figure 3 – Performance of the direct and AMGCL solvers for the (a) two- and (b) three-dimensional forward travel time test problems. (UMFPACK: Direct sparse solver from MATLAB; BiCGSTAB/ILU(0): Bi-conjugate gradient stabilized solver preconditioned with ILU(0) from MATLAB; BiCGSTAB/AMG/RS: Bi-conjugate gradient stabilized solver preconditioned with AMG using Ruge-Stüben coarsening; BiCGSTAB/AMG/SA: Bi-conjugate gradient stabilized solver preconditioned with AMG using smoothed aggregation coarsening).

Figure 4 – Comparison of the memory usage by the tested forward travel time solvers for the homogeneous three-dimensional benchmark problem.

Figure 5 – Groundwater (a) flow and (b) Transport speed up factors for the three-dimensional benchmark problem.

**Figure 1**

(a)

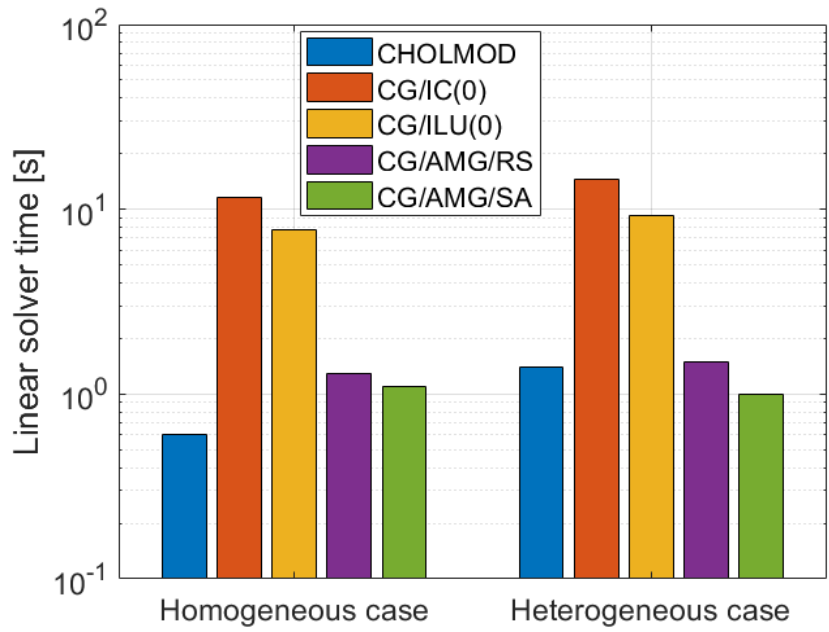

(b)

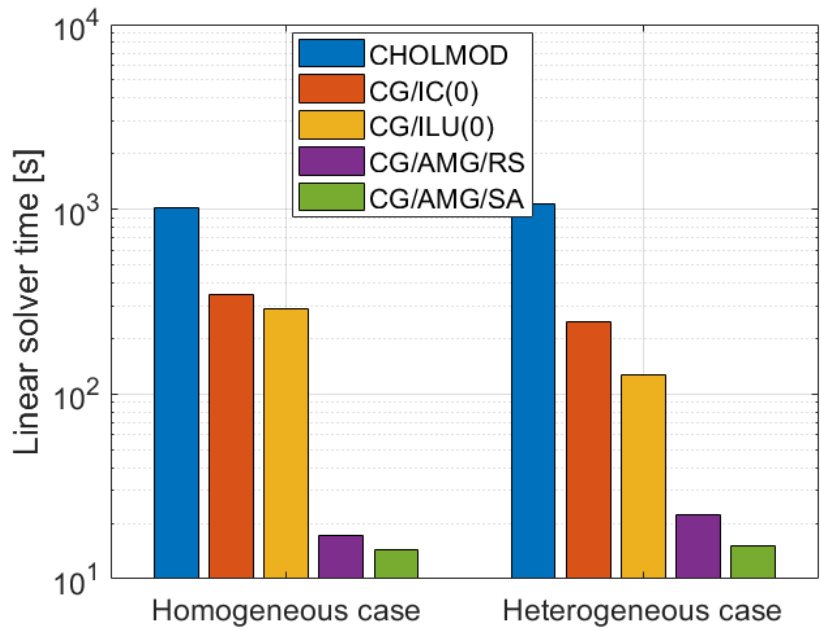

**Figure 2**

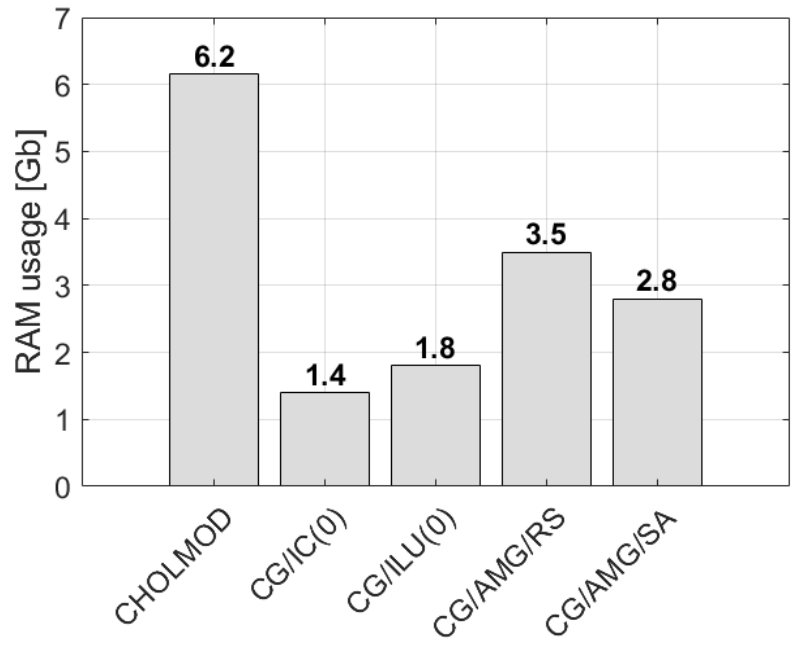

**Figure 3**

(a)

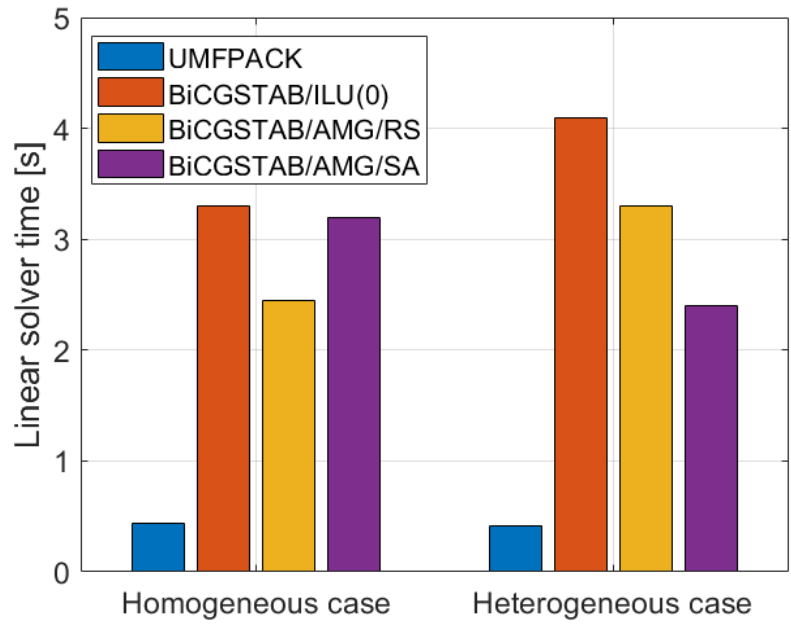

(b)

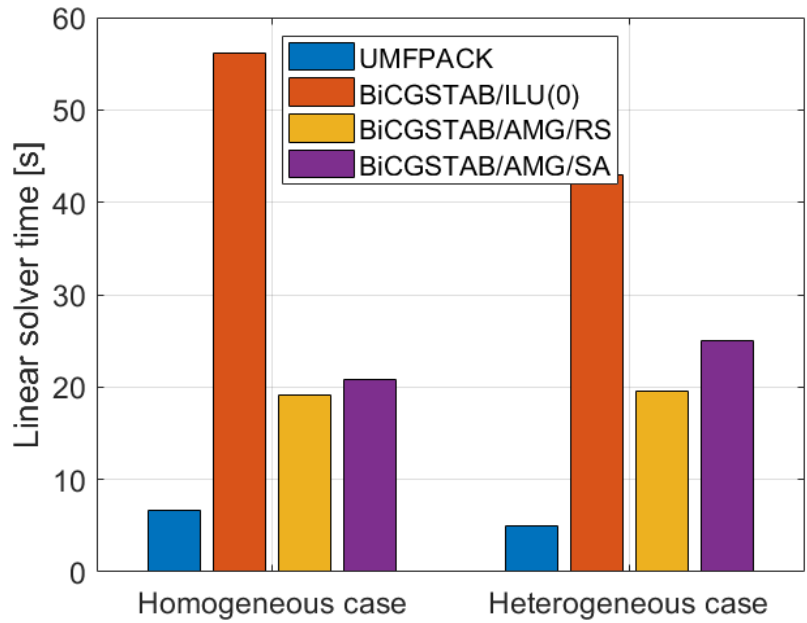

**Figure 4**

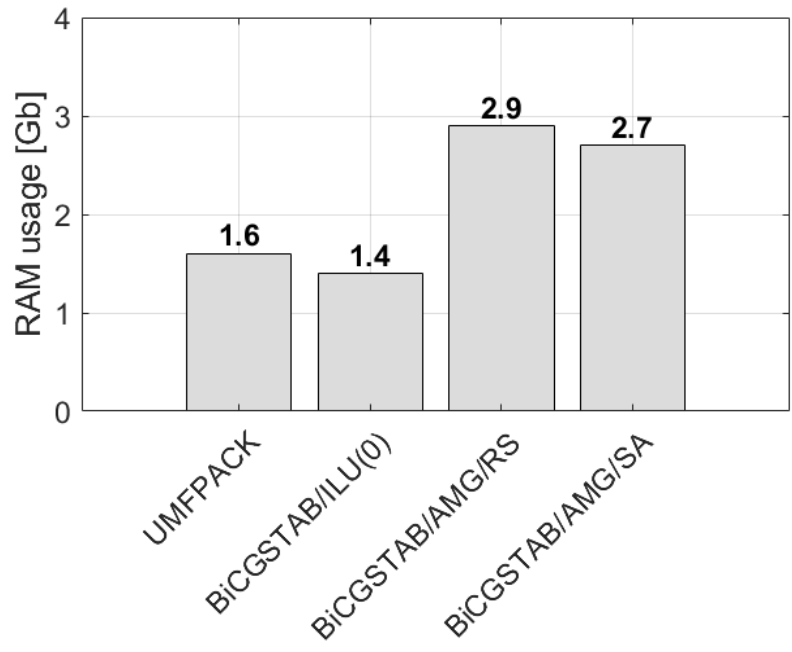

**Figure 5**

(a)

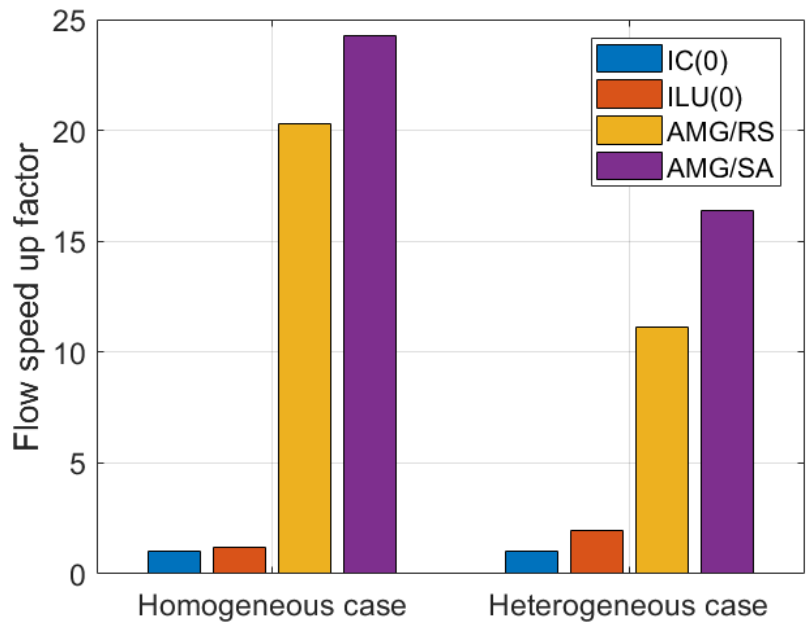

(b)

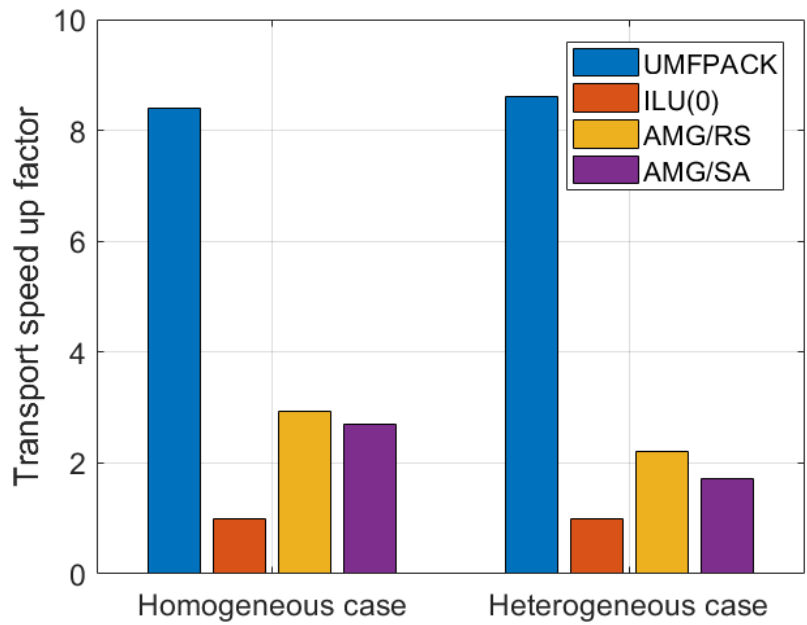